\documentclass[conference,10pt]{IEEEtran}

\usepackage[dvips]{color}
\usepackage{amsmath}
\usepackage{bm}

\usepackage[dvips]{graphicx}
\usepackage[english]{babel}
\usepackage[latin1]{inputenc}

\usepackage{amssymb}
\usepackage{url}
\usepackage{algorithmic}
\usepackage{algorithm}

\usepackage{cite}
\usepackage{stfloats} 

\begin{document}

\title{Comparison of SCIPUFF Plume Prediction with Particle Filter Assimilated Prediction for \\ Dipole Pride 26 Data}

\author{\IEEEauthorblockN{Gabriel Terejanu}
\IEEEauthorblockA{Department of Computer \\Science and Engineering\\
University at Buffalo\\
Buffalo, NY 14260\\
terejanu@buffalo.edu}
\and 
\IEEEauthorblockN{Yang Cheng}
\IEEEauthorblockA{Department of Mechanical \\and Aerospace Engineering \\
University at Buffalo \\
Buffalo, NY 14260 \\
cheng3@buffalo.edu}
\and 
\IEEEauthorblockN{Tarunraj Singh}
\IEEEauthorblockA{Department of Mechanical \\and Aerospace Engineering \\
University at Buffalo \\
Buffalo, NY 14260 \\
tsingh@eng.buffalo.edu}
\and 
\IEEEauthorblockN{Peter D. Scott}
\IEEEauthorblockA{Department of Computer \\Science and Engineering \\
University at Buffalo \\
Buffalo, NY 14260\\
peter@buffalo.edu}}

\maketitle

\selectlanguage{english}

\begin{abstract}
This paper presents the application of a particle filter for data assimilation in the context of puff-based dispersion models. Particle filters provide estimates of the higher moments, and are well suited for strongly nonlinear and/or non-Gaussian models. The Gaussian puff model SCIPUFF, is used in predicting the chemical concentration field after a chemical incident. This model is highly nonlinear and evolves with variable state dimension and, after sufficient time, high dimensionality. While the particle filter formalism naturally supports variable state dimensionality high dimensionality represents a challenge in selecting an adequate number of particles, especially for the Bootstrap version. We present an implementation of the Bootstrap particle filter and compare its performance with the SCIPUFF predictions. Both the model and the Particle Filter are evaluated on the Dipole Pride 26 experimental data. Since there is no available ground truth, the data has been divided in two sets: training and testing. We show that even with a modest number of particles, the Bootstrap particle filter provides better estimates of the concentration field compared with the process model, without excessive increase in computational complexity.
\end{abstract}


%
\IEEEpeerreviewmaketitle
\section{Introduction}
\label{sec:Introduction}
There is an increase in requirement for accuracy and computational performance in atmospheric transport and diffusion models used in critical decision making in the context of chemical, biological, radiological, and nuclear (CBRN) incidents. The output (field concentrations and dosages) of the dispersion models is used directly to guide decision-makers, and as an input for higher fusion levels, such as situation and threat assessment. Therefore the accuracy of the models as well as the time of delivery of the forecasts plays an important part in decision making.

Gaussian dispersion models have been extensively studied and used in assessing the impact of CBRN incidents. They have gained popularity due to their straightforward theoretical approach, and due to their relatively low computational complexity\cite{Arya1999}.

For accurate CBRN output, one cannot rely solely on mathematical models or on measurements recorded by the sensors on the field, because of their uncertainties. Thus, for better estimates and lower uncertainty, a fusion step, called Data Assimilation, is necessary to combine the model forecasts and the measurements.

In Data Assimilation, the estimation of the unknown system states given the underlying dynamics of the system and a set of observations may be framed as a filtering, smoothing or prediction problem. Given a fixed discrete time interval, $\{t_1,t_2, \ldots t_N\}$, over which observations are available, the problem of filtering is to find the best state at time $t_k$ given all the observations prior to and including $t_k$. The smoothing problem is to find the best state at time $t_k$ given all the observations up to time $t_N$, where $t_k \le t_N$. For $t_k > t_N$ the prediction problem is to forecast the state of the system at time $t_k$ using all the observations in the given interval.

For a linear system, under the assumption of Gaussian probability distributions, the problem of estimating the states of the system has an exact closed form solution given by the Kalman Filter\cite{Kalman1960}. If the  probability distributions are non-Gaussian or the system is nonlinear, in general no closed-form solution is available and different assumptions and approximations have been made for quasi-optimal solutions maintaining both accuracy and tractability. The systems considered here are in general nonlinear, but the assumption that the process and observation uncertainties can be adequately modeled as Gaussian is made.

The nonlinear filtering problem has been extensively studied and successfully employed in many applications, with various methods provided in the literature. Among the best understood and most frequently cited are the Extended Kalman Filter (EKF), the Ensemble Kalman Filter (EnKF), and more recently the Unscented Kalman Filter (UKF), and the Particle Filter (PF). The EKF is historically the first, and still the most widely adopted approach to nonlinear estimation problems. It is based on the assumption over small time increments, nonlinear system dynamics can be accurately modeled by a first-order Taylor series expansion\cite{Crassidis2004}. 

The PF uses a sampling approach, estimating the posterior probability distribution, including its higher order moments, by propagating and updating a number of particles, without the assumption of Gaussian statistics~\cite{Ristic2004}. The variable and high dimensionality of the state vector, which poses a problem to the standard nonlinear assimilation techniques, can be
dealt with using Particle Filters~\cite{Reddy2007}. Daum, et al.~\cite{Daum2003} showed
that a carefully designed Particle Filter should mitigate the curse of dimensionality for certain filtering problems.

This paper presents an implementation of the Bootstrap Particle Filter to assimilate chemical concentration readings of the Dipole Pride 26 experiment \cite{BiltoftJan1998} into the SCIPUFF dispersion model. The Particle Filter takes into account the uncertainty due to the data errors in the observed meteorology. The results show that the particle filter, with a modest number of particles, provides better estimates of the concentration field compared with the process model, underlying the importance of  meteorological data accuracy in CBRN incidents.

Data assimilation based on sampling techniques for CBRN incidents and numerical weather prediction (NWP), such as Particle Filter \cite{Reddy2007}, Unscented Kalman Filter \cite{Terejanu2007, Cheng2007} or Ensemble Kalman Filter \cite{Stuart2007, Anderson}, become more and more accessible as parallel computing becomes mainstream with the introduction of multi-core processors and multiple CPU computers. Sampling techniques such as Monte Carlo analysis \cite{Hanna2001,Bergin1999} or ensemble method \cite{Dabberdt2000,Draxler2003} have been used before to account for uncertainty due to data errors.

The Dipole Pride 26 (DP26) field experiment and the SCIPUFF dispersion model are presented in Section \ref{sec:DP26}. The Bootstrap Particle Filter is introduced in Section \ref{sec:PF} and its implementation for this particular problem is described in Section \ref{sec:implementation}. Numerical results on the CBRN scenario are presented in Section \ref{sec:results} and the conclusions and future work are discussed in Section \ref{sec:conclusion}.

\section{Dipole Pride 26 and SCIPUFF}
\label{sec:DP26}
The Dipole Pride 26 field experiment has been designed to validate transport and diffusion models at mesoscale distances \cite{BiltoftJan1998}. The experiment has been conducted at Yucca Flat, Nevada where gaseous sulfur hexafluoride (SF6) has been released in a series of instrumented trials, of which only $17$ provide useful puff dimension information. 

\begin{figure}
	\centering
		\includegraphics[width=3.4in]{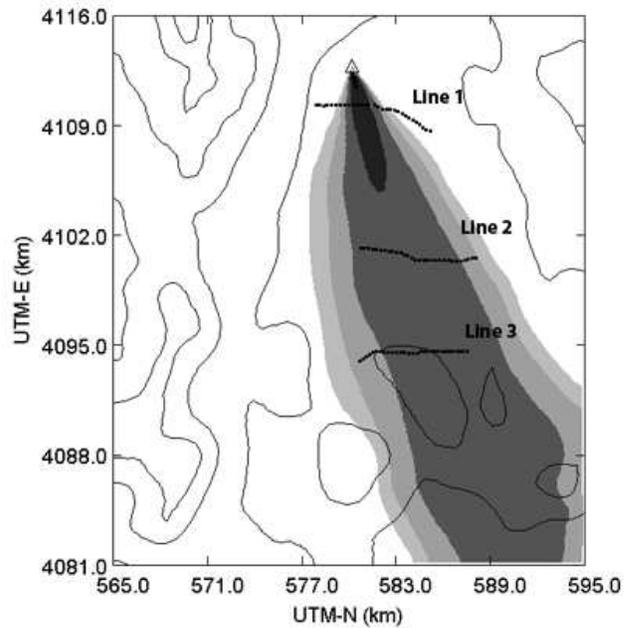}
	\caption{Process Model: chemical dosage plot after $3$hr at $1.5$m}
	\label{fig:experiment}
\end{figure}

Three lines of sensors, Fig.\ref{fig:experiment}, each with $30$ whole air-samplers have been used to record average concentrations every $15$~min. The chemical sensors, known also as sequential bag samplers ($12$ bags per sensor), are placed at $1.5$~m above the ground and spacing along lines is about $250$~m. The total sampling time of the chemical sensor is $3$~hr, hence total experimental duration for each trial to be monitored is $3.5$~hr. This is achieved  by delaying the acquisition of the last line of sensors with $30$~min. Six continuous SF6 analyzers, TGA-4000, have been used to record high-frequency variations of the gas concentration field, but their placement does not offer enough resolution to capture the crosswind structure of the chemical plume, and for the purpose of this paper these readings have been excluded from the study.

Eight Meteorological Data (MEDA) stations were used to provide surface-based meteorological measurements and two pilot balloon stations and one radiosonde provided the upper-air meteorological profiles. The meteorological measurements recorded provide information about the wind direction and speed, temperature, pressure and humidity. The variation of the wind field is given by the standard deviation of hourly wind speeds and directions recorded at the MEDA stations which are about $0.5-2$~m~s$^{-1}$ and $10^\circ-30^\circ$ respectively~\cite{Chang2003}.

In this paper we are focusing only of trial number six, which took place in Nov 12, 1996 and where a mass of $11.6$~kg of SF6 had been release from $6$~m height at the North dissemination site N2 as in Fig.\ref{fig:experiment}.
The dispersion model used to predict the chemical concentrations and dose at the sensor locations is SCIPUFF \cite{SykesJan2006}. SCIPUFF is a Lagrangian puff dispersion model that outputs the chemical concentrations at specified locations as cumulative contributions of Gaussian puffs. SCIPUFF is the dispersion engine incorporated into Hazard Prediction and Assessment Capability (HPAC) tool, used by the Defense Threat Reduction Agency (DTRA) for situation assessment of CBRN incidents. Besides the mean concentration field, SCIPUFF is also providing the uncertainty in the concentration value due to the stochastic nature of the turbulent diffusion process. The simulation is driven by the meteorological data, which in this case is provided by the MEDA stations, from which a wind field is created.

The observed wind data are fit in a least square sense, using variational methodology. An initial gridded wind field is constructed from the observation data by interpolation. Adjustments are then made to the initial 3D interpolated wind field vectors so as to satisfy conservation of mass in a way that also minimizes an integral function of the difference between the initial and adjusted fields. 

Existing literature provides evaluation studies of SCIPUFF with Dipole Pride 26 \cite{Chang2003,Cox2006,Chang2002}. The results reported show that SCIPUFF predictions are within a factor of $2$ of observations about $50\%$, where the model evaluation was based on maximum dosage anywhere on the sampling lines~\cite{Chang2003}. The studies emphasize the importance of wind field in the chemical concentration prediction accuracy and conclude that SCIPUFF prediction performance is comparable or better than other dispersion models.
\section{Boostrap Partcile Filter}
\label{sec:PF}
In this section we focus our attention on sequential state estimation using sequential Monte Carlo (SMC). SMC is known also as bootstrap filtering, particle filtering, the condensation algorithm, interacting particle approximations and survival of the fittest \cite{Arulampalam2002}.

Consider the following nonlinear system, described by the difference equation and the observation model:
\begin{eqnarray}
\textbf{x}_{k+1} &=& \textbf{f}(\textbf{x}_k) + \textbf{w}_k \label{process} \\
\textbf{z}_k &=& \textbf{h}(\textbf{x}_k) + \textbf{v}_k \label{measurement}
\end{eqnarray}

Denote by $\textbf{Z}_k = \{\textbf{z}_i | 1\leq i \leq k \}$ the set of all observations up to time $k$, conditionally independent given the process with distribution $p(\textbf{z}_k|\textbf{x}_k)$. Also, assume that the state sequence {$\textbf{x}_k$} is an unobserved (hidden) Markov process with initial distribution $p(\textbf{x}_0)$ and transition distribution $p(\textbf{x}_{k+1}|\textbf{x}_k)$.

Our aim is to estimate recursively the posterior distribution $p(\textbf{x}_k|\textbf{Z}_k)$ and expectations of the form \cite{Doucet}:
\begin{eqnarray}
E[\textbf{z}_k] = E[h(\textbf{x}_k)] = \int \mathbf{h}(\textbf{x}_k) p(\textbf{x}_k|\textbf{Z}_k) d\textbf{x}_k
\end{eqnarray}

In particle filters, the posterior distribution $p(\textbf{x}_k|\textbf{Z}_k)$ is approximated with $N$ weighted particles $\{{\bf x}_k^{(i)},w_k^{(i)}\}_{i=1}^N$, given~by
\begin{equation}
p(\textbf{x}_k|\textbf{Z}_k) \approx \sum_{i=1}^N w_k^{(i)}\delta_{\mathbf{x}_k^{(i)}}(\mathbf{x}_k)
\end{equation}

where ${\bf x}_k^{(i)}$ are the particles drawn from the importance function or proposal distribution, $w_k^{(i)}$ are the normalized importance weights that sum up to one and $\delta_{\mathbf{x}_k^{(i)}}(\mathbf{x}_k)$ denotes the Dirac-delta mass located in $\mathbf{x}_k^{(i)}$. Thus the expectation of a known function $\mathbf{h}(\mathbf{x}_k)$ with
respect to $p(\textbf{x}_k|\textbf{Z}_k)$ is then approximated by
\begin{equation}\label{}
    \int \mathbf{h}(\textbf{x}_k) p(\textbf{x}_k|\textbf{Z}_k) d\textbf{x}_k \approx \sum_{i=1}^N w_k^{(i)}\mathbf{h}(\mathbf{x}_k^{(i)})
\end{equation}

Suppose that we cannot sample from the posterior distribution and use an importance sampling approach to sample from a proposal distribution $q(\textbf{x}_k|\textbf{Z}_k)$. Hence, we can recursively update the weights:
\begin{eqnarray}
w_{k+1}^i &=& w_k^i \frac{p(\textbf{z}_{k+1}|\textbf{x}_{k+1}^i)p(\textbf{x}_{k+1}^i|\textbf{x}_k^i)}{q(\textbf{x}_{k+1}^i|\textbf{x}_k^i,\textbf{Z}_{k+1})} \label{weight1}
\end{eqnarray}

After a few iterations all but one particle will have negligible weights. Hence the algorithms allots time to update a large number of weights with no effect in our sampling, effect known as degeneracy problem. This problem can be overcome by adding a resampling strategy to Sequential Importance Sampling (SIS) Algorithm \ref{SIS}. 
\begin{algorithm}
\caption{Sequential Importance Sampling (SIS) algorithm}
\label{SIS}
\begin{algorithmic}[1]
\STATE Draw $\textbf{x}_k^i \sim q(\textbf{x}_k|\textbf{x}_{k-1}^i,\textbf{Z}_k)$ for $i=1 \ldots N$
\STATE Compute the importance weights \\ $w_{k+1}^i = w_k^i \frac{p(\textbf{z}_{k+1}|\textbf{x}_{k+1}^i)p(\textbf{x}_{k+1}^i|\textbf{x}_k^i)}{q(\textbf{x}_{k+1}^i|\textbf{x}_k^i,\textbf{Z}_{k+1})}$
\STATE Normalize the importance weights $\bar{w}_k^i = \frac{w_k^i}{\sum^{N}_{i=1} w_k^i}$
\STATE Multiply/Discard particles $\{\mathbf{x}_{k+1}^{(i)}\}_{i=1}^N$ with respect to high/low
importance weights $w_{k+1}^{(i)}$ to obtain $N$ new particles $\{\mathbf{x}_{k+1}^{(i)}\}_{i=1}^N$ with equal weights.
\end{algorithmic}
\end{algorithm}

The importance function plays a significant role in the particle filter. Usually, it is difficult to find a good proposal distribution especially in a high dimensional space. One may choose to approximate it using different methods, thus different flavor of particle filter. One of the simplest importance function is given by
\begin{eqnarray}
	q(\textbf{x}_{k+1}^i|\textbf{x}_k^i,\textbf{Z}_{k+1}) = p(\textbf{x}_{k+1}^i|\textbf{x}_k^i) \label{assum}
\end{eqnarray}

This implementation is called the Bootstrap Particle filter. By substituting \eqref{assum} back into \eqref{weight1} the new weight update equation becomes:
\begin{eqnarray}
w_{k+1}^i &\propto& w_k^i p(\textbf{z}_{k+1}|\textbf{x}_{k+1}^i) \label{weight2}
\end{eqnarray}

The resampling step reduces the sample impoverishment effect but introduces new practical problems: it limits the opportunity to have an efficient parallel algorithm, and particles with high weights are statistically selected many times and leads to a loss of diversity among the particles \cite{Arulampalam2002}.
\section{Particle Filter Implementation}
\label{sec:implementation}
Three types of uncertainties \cite{Rao2005} are present in the model predictions: \textbf{model uncertainty} due to the inaccurate representation of the chemical and dynamical processes, \textbf{data uncertainty} due to the errors in data used to drive the model and \textbf{random turbulence} of the atmosphere.

The model uncertainty is not completely reducible, and in the SCIPUFF prediction case it is estimated. The uncertainty due to the variability of the atmosphere cannot be further reduced and the uncertainty due to data errors it is usually high, more than $50\%$ of the total uncertainty \cite{Lewellen1989}, but it can be minimized. The data errors considered in this paper are meteorological data: wind speed and wind direction. The errors are due to the unrepresentative sitting of the wind sensors in the field and sensor accuracy and calibration \cite{Chang2002}. This information is rarely available and for this study a standard deviation of $0.5$~m~s$^{-1}$ for the wind speed and $5^\circ$ for the wind direction has been considered.

A FORTRAN implementation of the Particle Filter has been specially created for SCIPUFF to run in parallel on the cluster hosted at the Center for Computational Research at University at Buffalo. This implementation of the particle filter is designed to account for the uncertainty in the wind sensor readings while coping with the challenges of the data assimilation for the CBRN incidents using Gaussian puff models: variable dimensionality and high dimensionality \cite{Reddy2007}.

Compared with the maximum dosage approach \cite{Chang2003}, the evaluation method used in this paper is to compare the  predicted dosage after every $15$ min with the corresponding observed dosage. 

Since there is no ground truth, the samples provided in the DP26 have been divided to two sets: \textbf{the training set} used to perform the data assimilation, composed of Line 1 and Line 2 of sensors, and \textbf{the testing set}, Line 3 of sensors, used for performance evaluation.

Due to the spatial and temporal distance between the wind readings all the considered independent random variables described by a Gaussian distribution with mean given by the nominal sensor reading and uncertainty given by the standard deviation mentioned above. Hence the particles, each representing a SCIPUFF instance, are propagated using wind field generated from data sampled from this distribution. Each particle outputs a dose field $d_j^i$ which depends on the wind field; here $j$ is the sensor number, $i$ is the particle number and $k$ is the time step. The estimated dosage $\hat{d}_j$ is given by the following relation:
\begin{align}
\hat{d}_i(k) &= \sum^N_{i=1}{w_{k}^i d_j^i(k)}
\end{align}
Here $w_{k}^i$ is given by Eq.\eqref{weight2}. The conditional probability present in Eq.\eqref{weight2} is unknown since the authors could not find information regarding the uncertainty in the concentration readings for the whole air-samplers. The concentrations readings of the sensors have been assumed to be independent random variables due to the spatial and temporal distances. The likelihood have been approximated with a Gaussian function given by:
\begin{align}
p(d_j(k)|d_j^i(k)) &= \mathcal{N}\big(d_j^i(k)\big|d_j(k),v(k)\big)
\end{align}
Here $d_j(k)$ is the observed dosage at the $j^{\mathrm{th}}$ sensor at time $k$ and $v(k)$ is the sample variance of the difference between observed dosages and predicted ones over all the particles. Any predicted dosage less than $1$ ppt-hr (including zero) has been set to $1$ ppt-hr and all the observed dosages less than $10$ ppt-hr have been ignored. Hence the total number of dosage values to be compared is $47$.

\section{Numerical Results}
\label{sec:results}

Both the process model predictions, using nominal wind sensor readings, and the particle filter predictions with perturbed wind field have been compared against the observed dosages on the third line of sensors.

To evaluate the performance of the particle filter compared with the process model, we consider the following statistical measures \cite{Hanna1993}: \textbf{FB} - fractional bias, \textbf{MG} - geometric mean bias, \textbf{NMSE} - normalized mean square error, \textbf{VG} - geometric variance, \textbf{FAC2} - fraction of predictions within a factor 2 of observations and \textbf{FAC3} - fraction of predictions within a factor 3 of observations.
\begin{align}
	\mathbf{FB} &= \frac{\overline{D_o}-\overline{D_p}}{0.5(\overline{D_o}+\overline{D_p})} \\
	\mathbf{MG} &= \mathrm{exp}(\overline{\mathrm{ln}D_o}-\overline{\mathrm{ln}D_p}) \\
	\mathbf{NMSE} &= \frac{\overline{(D_o-D_p)^2}}{\overline{D_o}~\overline{D_p}} \\ 
	\mathbf{VG} &= \mathrm{exp}\left(\overline{(\mathrm{ln}D_o-\mathrm{ln}D_p)^2}\right) \\
	\mathbf{FAC2} &= \textrm{\#} D_p ~ \textrm{such that} ~ \frac{1}{2} \le \frac{D_p}{D_o} \le 2  \\
	\mathbf{FAC3} &= \textrm{\#} D_p ~ \textrm{such that} ~ \frac{1}{3} \le \frac{D_p}{D_o} \le 3	
\end{align}

Here $D_o$ represents the observed dosage and $D_p$ the predicted dosage. Since we are dealing with random samples, $50$ Monte Carlo runs have been performed in assessing the performance of the particle filter. The numerical results based on the performance metrics have been tabulated in Table \ref{results_table}. 

\begin{figure}
	\centering
		\includegraphics[width=3.4in]{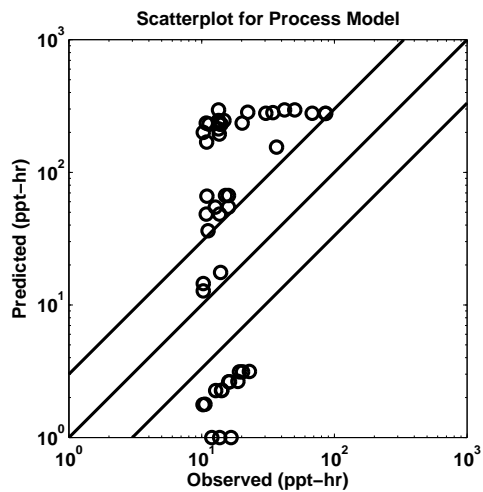}
	\caption{Process Model with nominal wind field}
	\label{fig:scatter_processmodel}
\end{figure}

\begin{figure}
	\centering
		\includegraphics[width=3.4in]{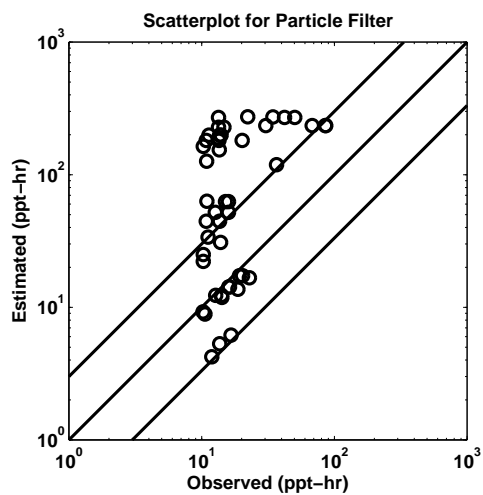}
	\caption{Particle Filter with perturbed wind field (best run)}
	\label{fig:scatter_particlefilter}
\end{figure}

Overall the particle filter provides improved estimates compared to the process model and all the performance measures are better on average. Since the predicted and observed values vary by several orders of magnitude, $\mathbf{MG}$, $\mathbf{VG}$, $\mathbf{FAC2}$ and $\mathbf{FAC3}$ are more appropriate.
\begin{table}[htbp]
\begin{center}
\renewcommand{\arraystretch}{2}
\caption{Numerical Results - $50$ Monte Carlo runs}
\label{results_table}
\begin{tabular}{|l|l|l|l|}
\hline
\ttfamily ~ & ~\textbf{Process Model}~ & ~\textbf{Particle Filter}~ & ~\textbf{PF 95\% CI}~ \\
\hline
~\textbf{FB}~ & ~$1.426$ & ~$\mathbf{1.405}$ & ~$1.390 - 1.419$ \\
\hline
~\textbf{MG}~ & ~$0.456$ & ~$\mathbf{0.443}$ & ~$0.402 - 0.484$ \\
\hline
~\textbf{NMSE}~ & ~$8.658$ & ~$\mathbf{8.466}$ & ~$8.249 - 8.683$ \\
\hline
~\textbf{VG}~ & ~$86.75$ & ~$\mathbf{61.20}$ & ~$53.65 - 68.74$ \\
\hline
~\textbf{FAC2}~ & ~$6.38\%$ & ~$\mathbf{11.45\%}$ & ~$9.31\% - 13.58\%$ \\
\hline
~\textbf{FAC3}~ & ~$6.38\%$ & ~$\mathbf{22.17\%}$ & ~$19.45\% - 24.89\%$ \\
\hline
\end{tabular}
\end{center}
\end{table} 
While we do not see a significant reduction in the geometric bias, the geometric variance and the fraction of factor $2$ and $3$ give a significant improvement of the particle filter over the process model. This is consistent with the scatter plots shown in Fig.\ref{fig:scatter_processmodel} and Fig.\ref{fig:scatter_particlefilter}. The particle filter is able to alleviate the under-prediction and over-prediction problem present in the dispersion models.

The result reiterates the need of accurate meteorological observations and provides support for the use of data assimilation in the CBRN incidents.

%


\section{Conclusion}
\label{sec:conclusion}
The paper presents an implementation of the Bootstrap Particle Filter to correct the SCIPUFF concentration predictions using concentration measurements provided by the chemical sensors deployed in the field. Due to the high uncertainty in the meteorological input, the CBRN dispersion models should be accompanied by a data assimilation step to account for this uncertainty and improve the predictions. For the Dipole Pride 26 the particle filter has doubled the number of predictions within a factor of $2$ of the observations and it has tripled the ones within a factor of $3$ of the observations.

A complete evaluation of the particle filter on all the Dipole Pride 26 trials and simulations with uncertainty for also the temperature, pressure and relative humidity, are planned as future work.
\section*{Acknowledgment}
\small
This work was supported by the Defense Threat Reduction Agency (DTRA) under Contract No. W911NF-06-C-0162. The authors gratefully acknowledge the support and constructive suggestions of Dr. John Hannan of DTRA. The authors would also like to thank Dr. Joseph C. Chang and Dr. Steve Hanna for their input regarding the Dipole Pride 26 dataset.

\bibliographystyle{IEEEtran}
\bibliography{CBD2008}

\end{document}